\documentclass[traditabstract,letter]{aa}
\usepackage{natbib}
\usepackage{graphicx}
\usepackage{txfonts}
\begin{document}
\title{Evidence of small-scale magnetic concentrations dragged by\\
vortex motion of solar photospheric plasma}
\subtitle{}

\author{L. Balmaceda\inst{1,2}, S. Vargas Dom\'inguez\inst{3}, J. Palacios\inst{1},  I. Cabello\inst{1} \and  V. Domingo\inst{1}
}

\offprints{L. Balmaceda, \email{laura.balmaceda@uv.es}
}

\institute{Image Processing Laboratory, University of Valencia, P.O. Box: 22085, E-46980 Paterna, Valencia, Spain
\and Instituto de Ciencias Astron\'omicas, de la Tierra y el Espacio, ICATE-CONICET, San Juan, Argentina
 \and Mullard Space Science Laboratory, University College London, Holmbury St Mary, Dorking, Surrey, RH5 6NT, UK
 }

\date{Received --; accepted 19 March 2010}

\abstract {Vortex-type motions have been {measured} by tracking bright points in high-resolution observations of the solar photosphere. These small-scale motions are thought to be determinant in the evolution of magnetic footpoints and their interaction with plasma and therefore likely to play a role in heating the upper solar atmosphere by twisting magnetic flux tubes. We report the observation of magnetic concentrations being dragged towards the center of a convective vortex motion in the solar photosphere from high-resolution ground-based and space-borne data.  We describe this event by analyzing a series of images at different solar atmospheric layers. By computing horizontal proper motions, we detect a vortex whose center appears to be the draining point for the magnetic concentrations detected in magnetograms and well-correlated with the locations of bright points seen in G-band and CN images.}


\authorrunning{Balmaceda et al.}

\titlerunning{Small-scale magnetic concentrations dragged by vortex plasma motion}

\maketitle

%
%________________________________________________________________

\section{Introduction}
Convection is an important mechanism of energy exchange responsible for the formation of the extensively studied solar granulation. The interaction between the convective plasma flows and solar magnetic fields causes the appearance of  solar structures on many spatial scales, sunspots and pores being the most conspicuous ones. Magnetism is often understood to be the main process controlling the whole  Sun's configuration and behavior, even in the so-called quiet regions where small magnetic elements such as bright points (hereafter BPs) are detected mainly along dark lanes in-between granules. Small-scale ($\lesssim$0.5~Mm) convectively driven vortex-type motions were discovered by \cite{bonet2008}.  {These motions were suggested by \citet{sturrock81} to account for about half of the photospheric kinetic energy, but this could not be confirmed due to the limitations of early observations. On larger spatial scales, i.e., granular sizes, vortical motions were observed by \citet{brandt1988}, \citet{tarbell91}, and \citet{title92}. Signatures of photospheric vortex flows have also been observed on even larger scales such as supergranular junctions \citep{attie2009}. Vortical motions in the solar photosphere were studied by simulations \citep[][and references herein]{zirker93} and are understood to play an important role in the evolution of magnetic footpoints and heating in the upper solar atmosphere \citep{brandt1988,wang1995}}. Here we describe a vortex motion in the solar photosphere that persists for about 20 minutes. It seems to affect the small-scale magnetic concentrations, which appear to be dragged towards the vortex center.

\begin{figure*}
\centering
\includegraphics[angle=-90,width=1.\linewidth]{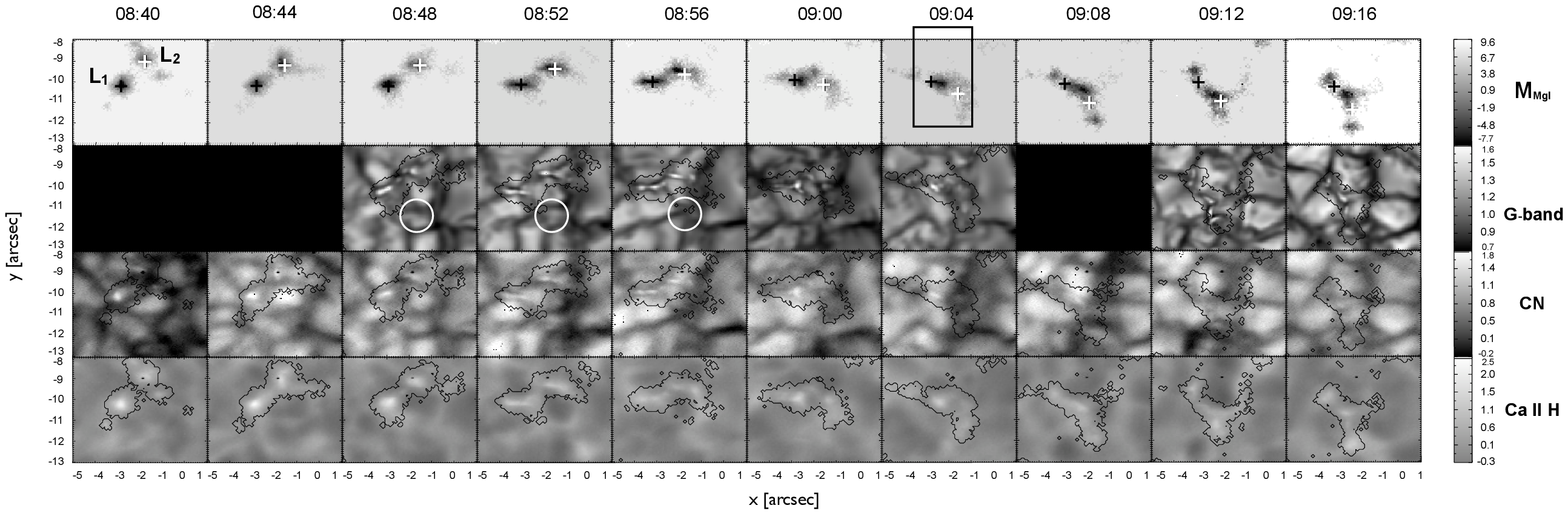} 
\caption{Quasi-simultaneous observations showing the evolution of a quiet sun region on 29 September 2007.  \emph{From top to bottom}: Mg I magnetogram, G-band, CN, Ca~II~H. Every panel displays the average image over 4-min intervals and the time stamps correspond to the initial time for each interval, respectively.  BPs are detected in all filtergrams and are co-spatial with the magnetic concentrations observed in magnetograms. Spatial scales in both axis are in arcsec and correspond to solar coordinates (FOV $\sim$6$\arcsec$$\times$5$\arcsec$). The black box at 09:04 UT shows the FOV covered by the SP in Fig.~\ref{lilia}. Encircled in white in G-band frames at 08:48, 08:52, and 08:56 UT is the location of the detected vortex (Sect.~\ref{S:3.3}) for reference.}
\label{secuencia}
\end{figure*}

\section{Observations and data processing}
\label{S:2}

An international campaign at the Canary Islands observatories was performed during September-October 2007 and corresponds to an important collaboration among several European and Japanese institutions.  Coordinated observations {involving all the} ground-based facilities at Canary Islands and the space solar telescope \emph{Hinode} \citep{kosugi2007} were performed as part of the \emph{Hinode Operation Program 14}.  In the present work, we use data recorded during a particular observing run on September 29, 2007, whose target was a quiet sun region north of the active region NOAA 10971,  close to the solar disk center ($\mu$=0.99).

\subsection{Ground-based observations} 

We used the  Swedish 1-m Solar Telescope \citep[SST,][]{SST} to acquire images in G-band ($\lambda 430.56$ nm) with an effective field-of-view (hereafter FOV) of 68$\farcs$54$\times$68$\farcs$54  and a sampling of $0.034$"/px. Image post-processing included the \emph{Multi-Object Multi-Frame Blind-Deconvolution} \citep[MOMFBD,][]{vannoort05} restoration technique to correct for atmospheric and instrumental aberrations. The final products were two time series of images (\emph{s1}: 08:47--09:07 UT and \emph{s2}: 09:14--09:46 UT) with a cadence of 15~s. Additional data post-processing steps were: compensation for diurnal field rotation, rigid alignment of the images, correction for distortion, and subsonic filtering to eliminate residual jittering \citep{title1986}.  

\subsection{Satellite data} 
The Solar Optical Telescope \citep[SOT,][]{tsuneta2008} onboard \emph{Hinode} acquired filtegrams in CN ($\lambda 388.3$ \emph{{\footnotesize FWHM}}~$0.8$ nm) and the core of Ca~II~H line ($\lambda 396.8 \pm 0.3$ nm) using the Broadband Filter Imager (BFI) {with a cadence of $\sim$35~s}. The observed FOV covered 19$\farcs$18 $\times$ 74$\farcs$09. {Although \emph{Hinode's} FOV is smaller than that of SST, they still overlap.}
Magnetograms in the Mg I line ($\lambda 517.3$ nm) were also obtained {at a cadence of $\sim$20~s} with the Narrowband Filter Imager (NFI).  Data from the spectropolarimeter \citep[SP,][]{ichimoto2008} were also acquired. This dataset consisted of Stokes parameters I, Q, U, and V measured along a slit of 256 pixels in raster scan mode of 18 scans from 08:20--09:44, with a sampling of 0.15$\arcsec$/px and FOV of $2\farcs66~\times~40\farcs57$. {The operation mode was set to \textit{dynamic mode}, i.e., the exposure time per slit position was 1.6~s to study the evolution of highly dynamic events. The noise level was $1.6 \times 10^{-3} I_c$ in Stokes I and $1.8 \times 10^{-3} I_c$ Stokes Q and U. }
The observed lines were the two FeI lines, $\lambda 630.15$ nm and $\lambda 630.25$ nm. The SOT images were corrected for dark current, flat-field, and cosmic rays  with standard procedures. A subsonic filter was applied to remove high frequency oscillations in filtergrams. SP data were inverted using the full atmosphere inversion code LILIA \citep{socasnavarro1997}, based on the SIR code \citep{ruizcobo1992}.

\begin{figure*}
\centering
\includegraphics[angle=90,width=1.\linewidth]{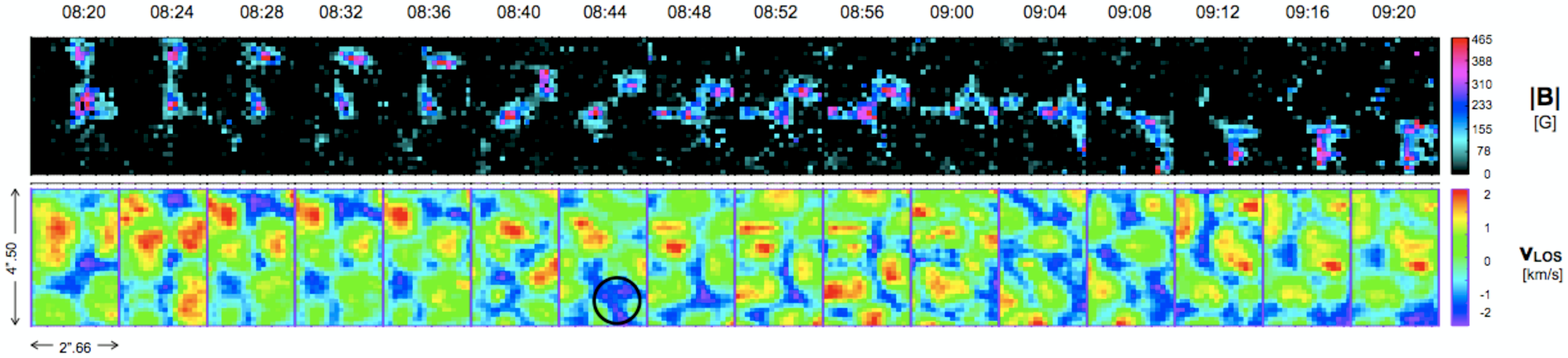} 
\caption{From top to bottom: magnetic field strength and LOS velocity single maps obtained by inverting SP data. The FOV covered is shown in the left lower panel. Positive/negative $V_{LOS}$ values correspond to upward/downward velocity directions. Encircled region in $V_{LOS}$ map at 08:44 UT displays a strong downflow appearing just before the formation of the vortex. Note that the displayed evolution of the region is longer in time here than in Fig.~\ref{secuencia} and the FOV smaller as framed by the black box at 09:04 UT in the same figure.}
\label{lilia}
\end{figure*}

\section{Data analysis and results}
\label{S:3}

\subsection{Multiwavelength description of the event} 
\label{S:3.1}

{Quasi-simultaneous} data taken by Hinode and SST were used. {The maximum time-lag between CN and Ca images was $\sim$2~s and  $\sim$12~s, respectively, between filtergrams and MgI magnetograms.}
Images were rotated, co-aligned, and trimmed to maintain the same FOV at all wavelengths.
Figure~\ref{secuencia} shows from bottom to top, time-averaged Ca II H, CN, and G-band filtergrams and Mg I magnetograms from 08:40 to 09:20 UT. Each frame corresponds to an average over 4 min, {obtained from $\sim$7 images in the case of filtergrams and $\sim$12 MgI magnetograms}. The averaging of the consecutive images allows us to easily identify the structures present in successive images. We superimpose on the corresponding filtergrams magnetogram contours, which are useful for visualizing the co-spatiality of the features observed at different wavelengths. The grayscale bars indicate the intensities normalized to the mean intensity of the entire data set for the respective wavelengths. The LOS magnetograms shown here consist of masks obtained after applying a threshold of 3$\sigma$ to remove the noise following \citet{krivova2004}. The good agreement between the magnetic features detected in these images and the bright elements present in the filtergrams gives us confidence to justify its use. {On first inspection, the magnetic feature appears to undergo an apparent rotation over the 40 minutes elapsed. We can identify a magnetic structure with two main lobes  (denoted by L1 and L2 in the first frame, top row of Fig.~\ref{secuencia}) one of which rotates around the other as indicated by the plus \emph{black} and \emph{white} plus-signs marking the 4-min averaged location of the centroid for L1 and L2, respectively\footnote{To identify the lobes, we use the line orthogonal to the linear fit of the pixels belonging to the structure and that passes through the center-of-mass of the whole data cube.}. }This movement is also detected at the various wavelengths, i.e., different heights in the solar atmosphere. Both G-band and CN data provide information about the photospheric level, while Ca II H images and Mg I magnetograms retrieve information from the low chromosphere. During the interval between 09:00 and 09:08, the L2 lobe seems to dissolve. {The magnetic flux density in this region} appears to decrease, but increases again when the whole structure has rotated and L2 approaches the lower part of the FOV. An increase in intensity is observed in Ca images and is accompanied by the appearance of BPs in G-band, while in CN the signal is more diffuse. As a consequence of the rotation, small-scale processes of fragmentation and coalescence of BPs clearly discernible in G-band filtergrams take place along the intergranular lane. The BPs are also detectable in CN images, although their appearance is not so well defined as in the SST data, especially towards the end of the sequence. The good agreement between the small elements detected in Ca II H and the other wavelengths is also remarkable. A forthcoming work (Vargas Dom\'inguez et al., 2010; hereafter Paper I) will focus on the evolution of  BPs in the sequence.
\begin{figure}
\centering
{\includegraphics[angle=-90,width=1.2\linewidth]{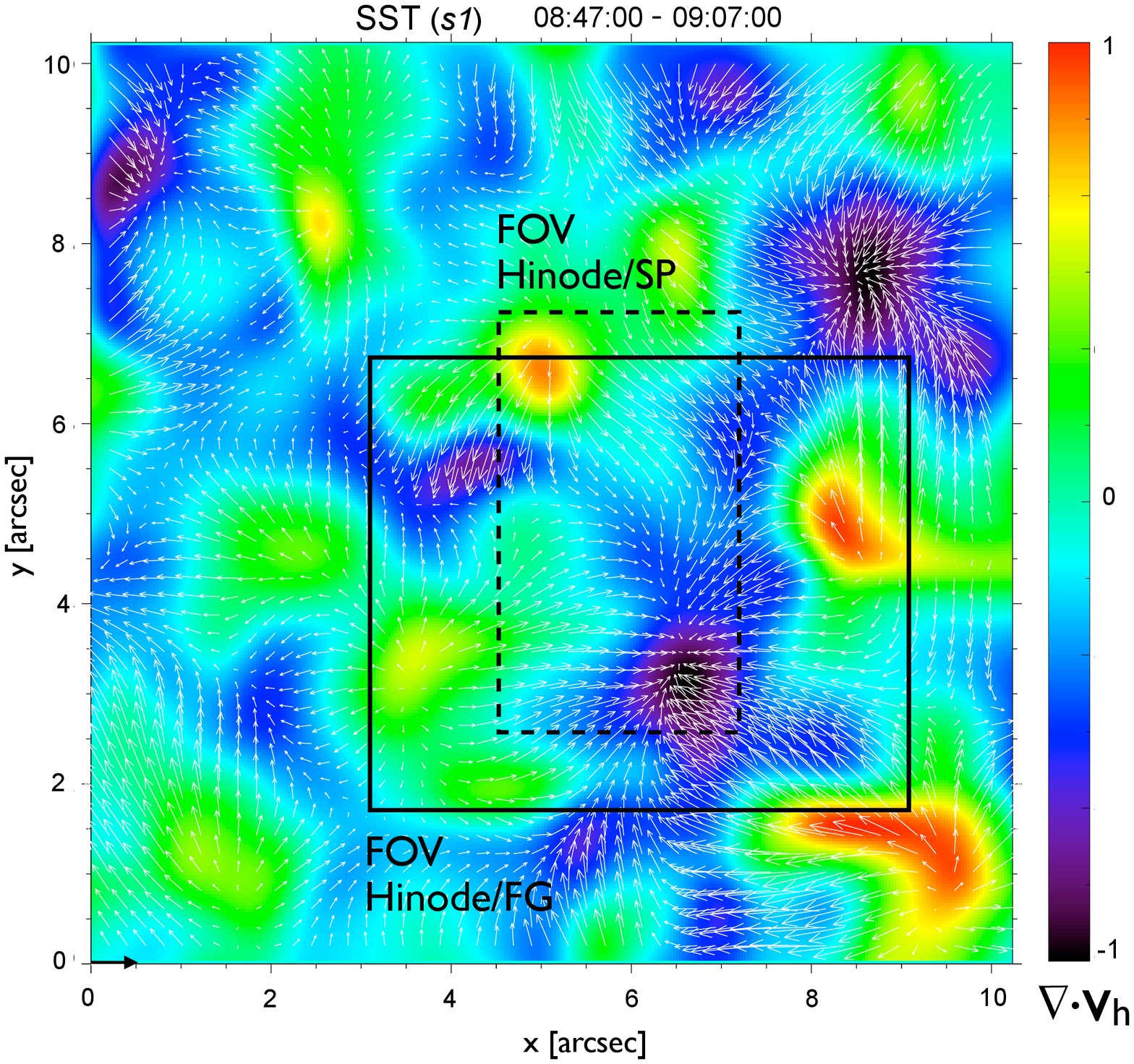}}
\caption{Map of horizontal velocities ({\footnotesize FWHM}~1\farcs0). The length of the black arrow at coordinates [0, 0] corresponds to  1.5 km  s$^{-1}$. The background image represents the normalized divergence field. The solid/dotted boxes extract the FOV covered by the sequences of images in Fig.~\ref{secuencia}/Fig.~\ref{lilia}, respectively, as labeled.}
\label{flowmap}
\end{figure}
\subsection{Characterization of the magnetic configuration of the region}
\label{S:3.2}

Figure~\ref{lilia} displays the evolution of the magnetic structure in the sequence of single maps of magnetic field strength $|B|$ retrieved by LILIA (first row) and LOS-velocity, $V_{LOS}$, {obtained by  estimating the Doppler shift in the line-center-wavelength using the center-of-gravity method to the Stokes \textit{I} profiles (second row).}
{For the inversion, only pixels with Stokes Q, U, or V amplitudes larger than 3 times their noise levels are taken into consideration to exclude profiles that cannot be inverted reliably.}
As observed in MgI magnetograms, the two magnetic lobes can be clearly distinguished in the $|B|$ maps displayed in the first row. SOT-SP data start earlier than the filtergrams and magnetograms described in the previous section. The first 5 frames display the evolution of the magnetic structure prior to 08:40, where we can see that it is already rotating. {The maximum value of the magnetic field strength was close to 600 G at 08:42, though not included in the displayed images.} 
{The rms of the errors in $|B|$ is 194~G. We defined the errors to be the differences between the inferred from the model and the real parameters at optical depth log $\tau$ = -1.}
{To check the reliability of the inversion, we also compared these values with those obtained by assuming a weak-field approximation \citep{landi92} and achieved similar results.}
{The longitudinal magnetic flux decreases by about 30\% during the whole time interval.} The area of the structure seems to shrink with time. In particular, for 09:00--09:20 the magnetic patches evolve from being bound together, when the two structures lose their identity and become elongated. The Y-shape structure that is evident in the MgI magnetogram (at about 09:12 UT) is not visible here. {One possible explanation is that Fe and MgI lines provide information about different heights in the solar atmosphere. The missing component of ``Y'' exhibits in MgI magnetograms a very weak and variable signal that is co-spatial with bright regions in CaII images. However, the component is absent in photospheric filtergrams and SP maps.} SP maps are available until 09:44 (though not shown in the figure) and show that the structure continues rotating after 09:20 when the SOT-NFI and BFI time series end, L1 and L2 approaching each other towards the end of the observing interval. The inclination (not shown here) remains stable during the whole observing period. Magnetic flux tubes are vertically oriented with inclination values close to 180 degrees. However, changes in the inclination can be seen at 08:44, when some BPs appear in-between the magnetic lobes. {In spite of the high noise level in Stokes Q and U, the good agreement between the magnetic structure observed in the LOS magnetogram and that of SP maps is indicative of nearly vertical magnetic fields.} In the $V_{LOS}$ maps (second row of Fig.~\ref{lilia}), the granules are clearly distinguished as upflows with velocities from 1 to 2 km  s$^{-1}$. The intergranular lanes are the areas showing downflows from -1.3 to -2 km  s$^{-1}$. A strong downflow region appears at 08:44 UT (encircled area in Fig.~\ref{lilia}) that denotes the location where the photospheric vortex forms (Sect.~\ref{S:3.3}). It remains there for several minutes (until 08:56) and at 9:20 another downflow region is observed at the location of L2 at the end of the sequence, though much less intense. The magnetic patches are located in the lanes, with a couple of exceptions where the magnetic areas do excursions into granules (at 09:00 and 09:07 UT). During the whole sequence, numerous small-scale processes take place leading to the formation of BPs (Paper I).

\subsection{Photospheric plasma flows}
\label{S:3.3}

\begin{figure}
\centering
\bigskip
\hspace{-8mm}
\includegraphics[angle=90,width=.9\linewidth]{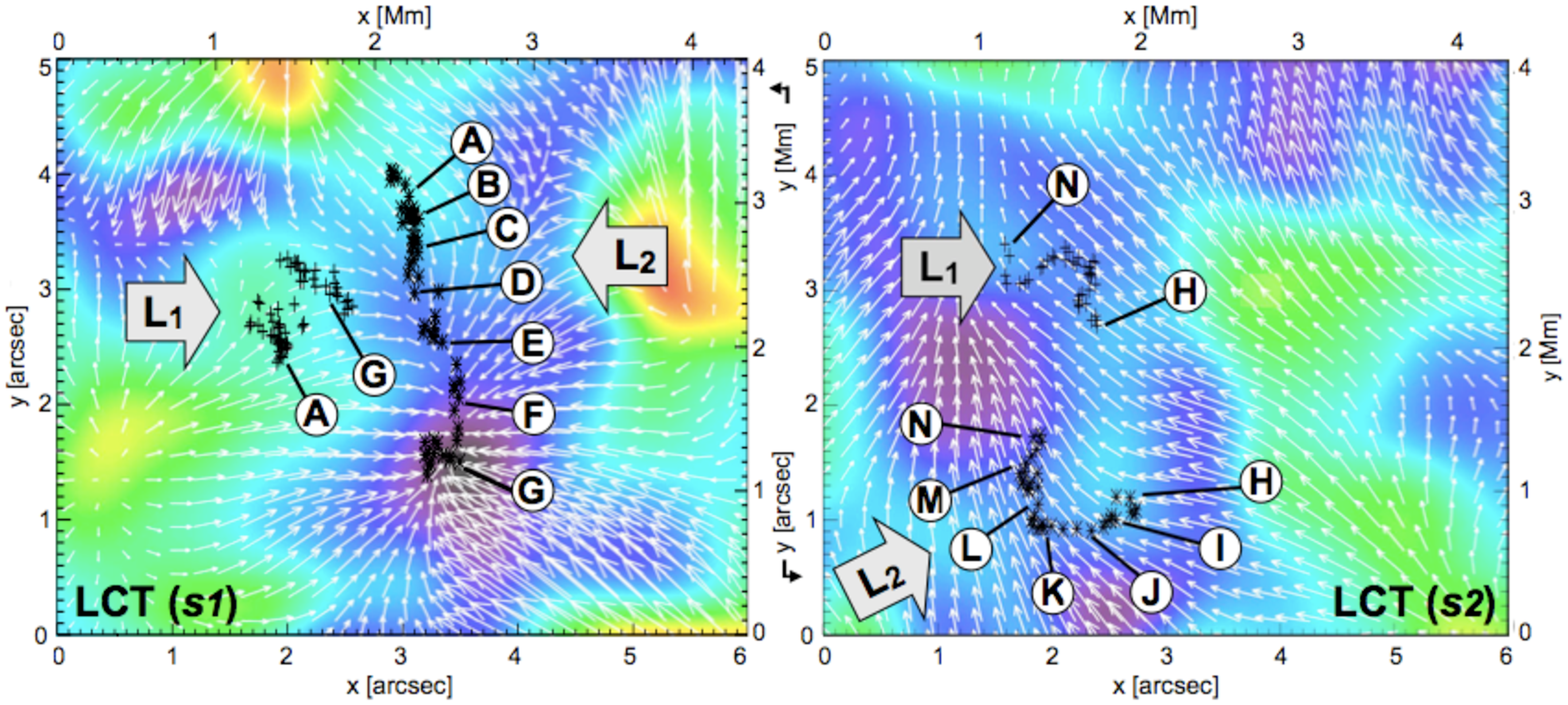} 
\caption{Temporal evolution of the magnetic centroid for both lobes ($L_1$ and $L_2$)  in Fig.~\ref{secuencia}. The trajectories followed by the centroids are plotted independently and superimposed on the flow maps computed for SST/G-band images in time intervals: \emph{s1} (\emph{left}) and \emph{s2}  (\emph{right}). The background false-color image is the divergence field computed from each respective flow map. Alphabet letters correspond to UT times as follows:  {\bf A}-08:40, {\bf B}-08:48, {\bf C}-08:56,  {\bf D}-09:00, {\bf E}-09:04, {\bf F}-09:08, {\bf G}-09:12, {\bf H}-09:21, {\bf I}-09:24, {\bf J}-09:28, {\bf K}-09:32, {\bf L}-09:36, {\bf M}-09:40,{\bf N}-09:44. The FOV is the same boxed in solid line in Fig.~\ref{flowmap}.}
\label{barycenter}
\end{figure}

The G-band series from SST were used to analyze the horizontal proper motions of structures in the FOV. Proper motions were measured by using the local correlation tracking technique \citep[LCT,][]{november1988} implemented by \cite{molowny1994}. Maps of horizontal velocities are calculated for the time series by using a Gaussian tracking window of {\footnotesize FWHM}~$1\farcs0$, i.e., roughly half of the typical granular size. Figure~\ref{flowmap} shows the flow map computed from the first of the SST time series (\emph{s1}). This map of horizontal velocities is calculated by averaging over the total duration of the series ($\sim20$~min). {The background image represents in false-color the normalized (factor of 0.086 s$^{-1}$) divergence field.}
Horizontal velocity magnitudes are averaging-dependent  values \citep{vargasthesis}, though the flow patterns yielded by the analysis of proper motions are commonly assumed to represent plasma motions thus enabling the detection of general trends, which is our main aim here (e.g., sinks or granulation downdrafts where the cold plasma returns to the solar interior). The map of horizontal displacements is dominated by flows coming from granular explosive events and commonly associated with mesogranulation \citep{roudier2004,bonet2005}. We identified only a pair of examples where strong sinks are present at intergranular lanes (coordinates [6.5, 3.5] and [8.5, 7.5] in Fig.~\ref{flowmap}) but, in contrast to other areas displaying downflows, these are characterized by being the draining point where all horizontal velocity vectors in the neighborhood converge. {This behavior was observed in numerical simulations by \citet{nordlund85}, who showed that the plasma on granular scales in strong downflow regions tend to rotate around the center, like a \emph{bathtub} effect.} Both vortex-type events last for several minutes ($>20$ min). Nevertheless, the one framed in the two boxes in Fig.~\ref{flowmap} is not only less stable and has a defined onset in our data (at $\sim$08:48 UT, Paper I) but also appears to be the final destination of the magnetic concentrations being dragged from an upper location (at coordinates [5, 7] in the same figure), as described in Sect.~\ref{S:3.1}.  The latter vortex remains rather stable but disappears abruptly in the map of horizontal velocities computed for \emph{s2}, which is dominated by an organized flow coming from the lower right part of the FOV (Fig.~\ref{barycenter},  right). Unfortunately, as already mentioned, there is a 7-min gap between \emph{s1} and \emph{s2} meaning that it is not possible to define exactly when and how the vortex disappears.

\section{Discussion}
\label{S:4} 

In Fig.~\ref{barycenter}, we plot the trajectories followed by the \emph{centroid} of the distribution of {magnetic flux density} for lobes L1 and L2. 
The \emph{centroids} denoted with plus-signs and asterisks for the respective lobes are now calculated for the individual maps obtained from SP data with a cadence of $\sim$36~s using the magnetic flux density $M$ for weighting. They are defined by $x_{c}=\sum_{i=1}^{N} x_{i} M(x_{i},y_{i}) / \sum M(x_{i},y_{i})$ and $y_{c}=\sum_{i=1}^{N} y_{i} M(x_{i},y_{i}) / \sum M(x_{i},y_{i})$, where ($x{i}$, $y_{i}$) are the coordinates of the $i$th pixel for each lobe. In spite of its lower resolution, we use the SP dataset because its time series extends until 09:44 UT, allowing us to track the dynamics of L1 and L2 over a longer period. The trajectories are superimposed on both the respective flow maps computed for the SST time series (\emph{s1} and \emph{s2}) and the divergence field for each series. {The comparison with the centroid positions calculated using the MgI magnetograms (top row in Fig.~\ref{secuencia}) allowed us to estimate the errors to be $\epsilon_{x1} = \pm0.35'' $, $\epsilon_{y1} = \pm0.38''$ for L1 and $\epsilon_{x2} = \pm0.32''$, $\epsilon_{y2} = \pm0.28''$ for L2.} {From the centroid positions, it was possible to estimate the horizontal velocities for each lobe}. While L1 motion is confined to a radius of approximately 1\arcsec, L2 travels a much larger distance. In the time interval prior to 09:14, its motion seems to be strongly influenced by the vortex formed in the coordinates [3.5, 1.5] in the left panel of Fig.~\ref{barycenter}, increasing its velocity {from $\sim$1.5 to 3~km~s$^{-1}$} as it approaches this location. After the vortex has disappeared, L2 continues its movement {with decreasing velocity}  probably influenced by the continuous evolution of the surrounding granules towards a new sink denoted by the region of negative divergence whose center is located at coordinates [1.0, 2.5] in Fig.~\ref{barycenter} (right). {Because of the lack of space here, additional analysis of centroids velocities will be deferred to Paper~I. }

{A possible explanation of the observed magnetic concentrations being dragged towards the vortex center could be that this is the result of the stochastic evolution of granules that allow the BPs to approach the vortex influence, increase their velocities, and eventually \emph{fall} into the vortex  (\emph{surface scenario}).  On the other hand, long-lasting ($>1$ h) vertical magnetic flux tubes anchored beneath the surface might be directly influenced by the vortex action due to some interaction mechanism taking place underneath. Although presenting additional evidence to support this scenario is beyond the scope of this paper, this possibility should not be ruled out.}

\begin{acknowledgements}
The Swedish 1-m  Solar Telescope is operated by the Institute of Solar Physics of the Royal Swedish Academy of Sciences at the Spanish Observatorio del Roque de los Muchachos of the Instituto de Astrof\'isica de Canarias. \emph{Hinode} is a Japanese mission developed and launched by ISAS/JAXA, with NAOJ as domestic partner and NASA and STFC (UK) as international partners. It is operated by these agencies in co-operation with ESA and NSC (Norway). The authors would like to thank the anonymous referee for helpful comments on the manuscript.
\end{acknowledgements}

\bibliographystyle{aa}

\end{document}